\title{Modulated magnetic structure in $^{57}$Fe doped orthorhombic YbMnO$_3$: a M\"ossbauer study}

\author{\textsc{Mathieu Duttine, Alain Wattiaux, Felix Balima} \\ CNRS, Universit\'e de Bordeaux, ICMCB, UPR 9048 \\ 33600 Pessac, France \\ \\
\textsc{Claudia Decorse} \\ ICMMO, Universit\'e Paris-Sud, Universit\'e Paris-Saclay \\ 91405 Orsay, France \\ \\
\textsc{Hicham Moutaabbid} \\ IMPMC, Sorbonne Universit\'es - UPMC, CNRS, IRD, MNHN \\ 75005 Paris, France \\ \\
\textsc{Pierre Bonville}\thanks{Contact author: pierre.bonville@cea.fr} \\ SPEC, CEA, CNRS, Universit\'e Paris-Saclay, CEA-Saclay \\ 91191 Gif-sur-Yvette, France}
\date{\today}

\documentclass[a4paper,11pt]{article}
\usepackage{graphicx}
\usepackage{amssymb}
\usepackage{lineno}

\begin{document}
\maketitle

\begin{abstract}
In the orthorhombic manganites o-RMnO$_3$, where R is a heavy rare earth (R = Gd-Yb), the Mn$^{3+}$ sublattice is known to undergo two magnetic transitions. The low temperature phase has an antiferromagnetic structure (collinear or elliptical), which has been well characterized by neutron diffraction in most of these compounds. The intermediate phase, occurring in a narrow temperature range (a few K), is documented for R = Gd-Ho as a collinear modulated structure, incommensurate with the lattice spacings. We report here on a $^{57}$Fe M\"ossbauer study of 2\% $^{57}$Fe doped o-YbMnO$_3$, where the spin only Fe$^{3+}$ ion plays the role of a magnetic probe. From the analysis of the shape of the magnetic hyperfine M\"ossbauer spectra, we show that the magnetic structure of the intermediate phase in o-YbMnO$_3$ (38.0\,K $<$ T $<$ 41.5\,K) is also modulated and incommensurate.
\end{abstract}

\vspace{2pc}
\noindent{\it Keywords}:
Modulated magnetic structures, Orthorhombic manganites,  M\"ossbauer spectroscopy

\section{Introduction} \label{intr}

The physics of the orthorhombic (or perovskite) rare earth manganites o-RMnO$_3$, where R is a rare earth ion, is governed by the interplay of various interactions: the largest is the Jahn-Teller effect on the $3d^4$ Mn$^{3+}$ ion, which can lead to orbital ordering; the interionic exchange interaction, which leads to magnetic ordering; and the magneto-electric coupling which couples ferroelectric order and magnetic order in certain given circumstances (for a review, see Ref.\cite{shuai}). Due to the present interest in multiferroic phenomena, i.e. acting on magnetic moments {\it via} an electric field (or vice-versa) \cite{kimura0}, the precise determination of the lattice and magnetic properties of these interesting materials, and of their interplay, is of prime importance.

The magnetic phase diagram of the orthorhombic rare-earth manganates has been thoroughly established in Ref.\cite{tachibana} by means of specific heat measurements. For the naturally occurring perovskite phases (R = La-Gd), a single magnetic transition, due to ordering of the Mn moments, is observed. Its critical temperature T$_{\rm N1}$ decreases with the rare earth radius. For Eu and Gd however, a second transition occurs at T$_{\rm N2} > \ $T$_{\rm N1}$. The same happens for all the heavier rare earth manganates, for which the orthorhombic phase is metastable. It is obtained through high pressure annealing of the naturally occurring hexagonal phase. Regarding the ground magnetic structure of o-RMnO$_3$, neutron diffraction studies determined it to be of antiferromagnetic (AF) collinear A-type for R=La-Gd \cite{goto, kimura}, of transverse spiral type for R = Tb and Dy  \cite{kenzelmann,arima} and of AF collinear E-type for R = Ho \cite{munoz}. In TbMnO$_3$ and DyMnO$_3$, the spiral magnetic order is accompanied by a ferroelectric order and a large magneto-electric effect \cite{kimura0,goto}. In HoMnO$_3$, the E-type order also bears ferroelectricity \cite{lorenz}. The structure of the intermediate phase (T$_{\rm N1} < {\rm T} < {\rm T}_{\rm N2}$) is documented only for R = Gd-Ho, where it was shown to be incommensurate sine-wave modulated \cite{munoz}. The N\'eel temperature T$_{\rm N1}$ characterizes therefore a lock-in transition below which the magnetic structure becomes commensurate with the lattice spacing. 

For the last rare earths of the series, the ground magnetic structure has been determined only for R = Yb \cite{huang}: it is of AF E-type, but the structure of the intermediate phase has not yet been elucidated. By analogy with what occurs for R = Gd-Ho, it is expected to be incommensurate modulated. A M\"ossbauer spectroscopy study of $^{57}$Fe doped o-YbMnO$_3$, with the isotopes $^{57}$Fe and $^{170}$Yb, was performed in Ref.\cite{stewart}, but could not reach a definitive conclusion about the structure of the intermediate phase. 

We have performed a $^{57}$Fe M\"ossbauer investigation of this material, doped at a 2\% level with $^{57}$Fe, and by analyzing the shapes of the spectra in the intermediate phase, we show they are compatible with a collinear incommensurate modulated type. Recently, we have shown the feasability of this method by analyzing the incommensurate magnetic phases in FeVO$_4$ by M\"ossbauer spectroscopy \cite{colson} (see also Ref.\cite{sobolev}).

\section{Sample preparation and magnetic measurements}

\subsection{Sample preparation and characterisation}

The hexagonal polymorph h-YbMn$_{0.98}$Fe$_{0.02}$O$_3$ was prepared as polycrystaline powder by solid state reaction. Oxides Yb$_2$O$_3$, MnO$_2$ and Fe$_2$O$_3$ of at least 99.99\% purity were used as precursors. The iron oxide was enriched in the $^{57}$Fe isotope up to 95.5\%.

\begin{figure}[!ht]
\centering\includegraphics[width=0.6\linewidth]{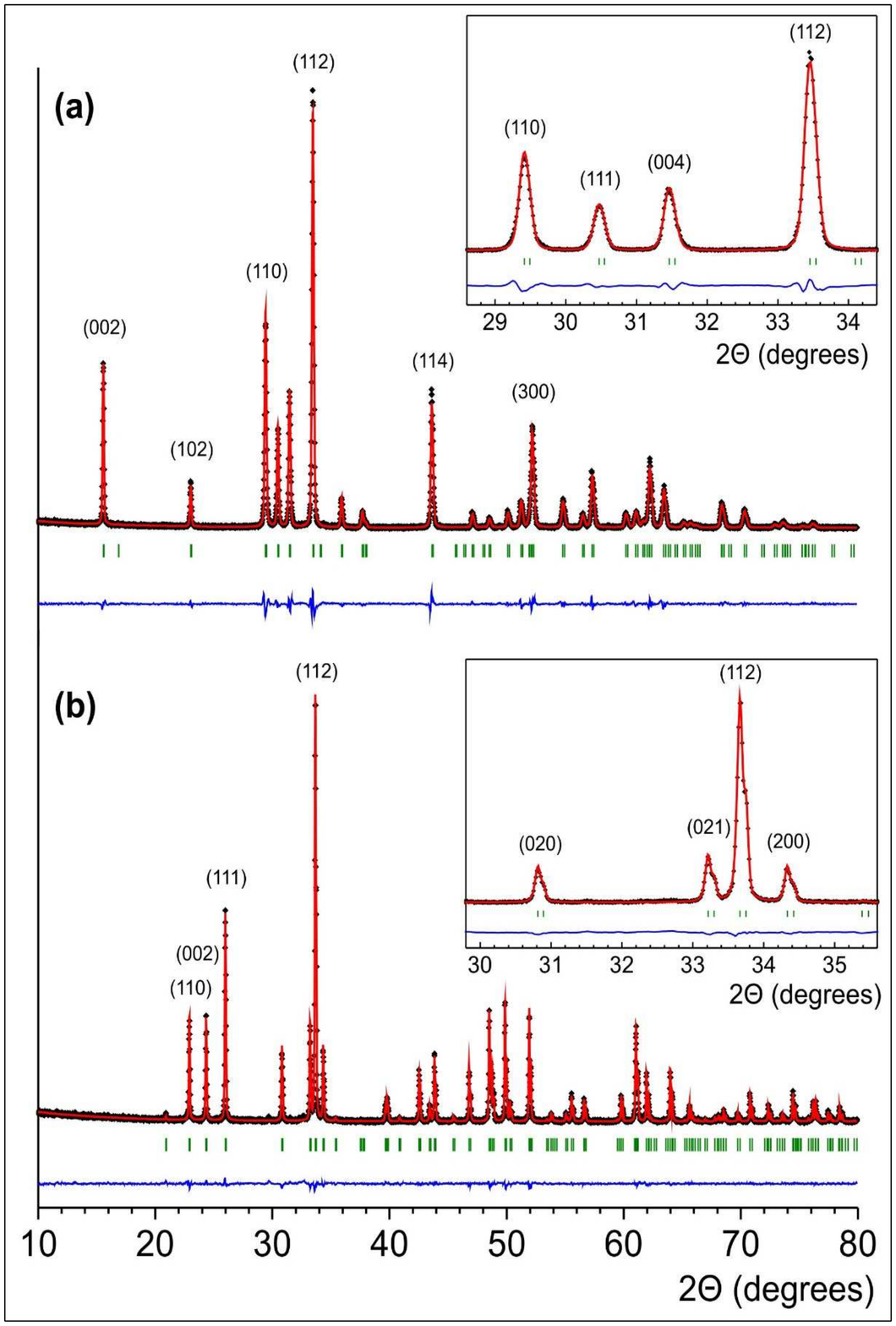}
\caption{Full-pattern matching (Le Bail) analysis of XRD patterns (Cu-K$\alpha_{\rm 1}$K$\alpha_{\rm 2}$) in 2\% $^{57}$Fe-doped h-YbMnO$_3$ ({\bf a}) and o-YbMnO$_3$ ({\bf b}). Black dots: experimental data, red line: calculated pattern, blue line: arithmetic difference between observed and calculated data, green marks: Bragg positions.}
\label{xrd}
\end{figure}
Stoichiometric quantities of the starting materials were thoroughly mixed and ground together, then pressed into pellets and heated up to 950$^\circ$C, in air, for 12h. After an intermediate grinding, the mixture was pressed again into pellets and further heated at 1100$^\circ$C for 48h in air, which completed the synthesis. According to the powder x-ray diffraction pattern (see Fig.\ref{xrd} {\bf a}), the h-YbMnO$_3$ sample was single phase, with $a$=$b$=6.068(1)\,\AA\ and $c$=11.364(1)\,\AA.

The orthorhombic phase of YbMnO$_3$ was prepared by heating the hexagonal phase h-YbMnO$_3$ under high pressure. The powder was packed into a platinum capsule surrounded by pyrophillite (a good pressure transmitter and thermal insulator) and heated at 1100$^\circ$ (ramp rate 40$^\circ$/min) during 40\,min under an applied pressure of 5\,GPa. The XRD pattern is entirely compatible with the orthorhombic Pnma space group (see Fig.\ref{xrd} {\bf b}), with lattice parameters (with the Pnma setting) : $a$ = 0.57988\,nm, $b$ = 0.73094\,nm and $c$ = 0.52197\,nm, in good agreement with published values \cite{huang,stewart}. The Bragg peaks are narrow, showing negligible nonstoichiometry.

\subsection{Magnetic susceptibility}

\begin{figure}[!ht]
\centering\includegraphics[width=0.8\linewidth,trim = 0 0 0 15cm]{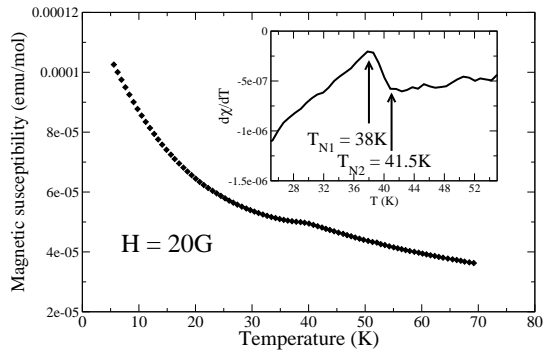}
\caption{Thermal variation of the magnetic susceptibility $\chi(T)$ in o-YbMnO$_3$ doped with 2\% $^{57}$Fe, in a field of 20\,G. The inset shows the derivative of $\chi(T)$.}
\label{xit}
\end{figure}

The polycrystal magnetic susceptibility was measured with a field of 20\,G in the temperature range 5 - 70\,K. It shows a monotonic decrease as temperature rises, with a tiny accident around 40\,K (see Fig.\ref{xit}). 

The structure of the accident is better revealed by the derivative $d\chi/dT$ shown in the inset of Fig.\ref{xit}. As temperature is decreased, one observes a jump at 41.5\,K, which is therefore identified as T$_{\rm N2}$, the transition temperature to the intermediate phase, and a maximum at 38\,K, which is identified as T$_{\rm N1}$, the transition temperature to the collinear E-type AF phase. Thus, in our o-YbMnO$_3$ sample, the intermediate magnetic phase occurs in the range 38.0\,K $<$ T $<$ 41.5\,K.
These boundaries match rather well the T$_N$ value of 43\,K reported in Ref.\cite{huang} and the two values T$_{\rm N1} \simeq$36\,K and T$_{\rm N2} \simeq$ 40\,K inferred from Ref.\cite{tachibana}, showing that the intermediate phase transition temperatures vary slightly from sample to sample.

\section{M\"ossbauer measurements}

The M\"ossbauer spectra were recorded with a linear velocity electromagnetic drive on which is mounted a commercial $^{57}$Co$^*$:Rh $\gamma$-ray source. A standard liquid He cryostat with temperature regulation was used.

\subsection{The Electric Field Gradient tensor}

o-YbMnO$_3$ crystallizes into the orthorhombic space group Pnma, where the Mn 4b-site has triclinic C$_i$ point symmetry. Fe substitutes for Mn in the doped material, and since Fe$^{3+}$ and Mn$^{3+}$ have approximately the same radius (0.645\,\AA), one can reasonably assume that the Fe site is not appreciably distorted with respect to the Mn site. On this basis, the authors of Ref.\cite{stewart} have performed a Point Charge Model calculation using the o-YbMnO$_3$ crystal parameters in order to obtain the Electric Field Gradient (EFG) tensor $V_{ij}$ (principal axes and diagonal values) at the Fe/Mn site. This tensor is needed in order to evaluate the quadrupole hyperfine interaction which characterizes the M\"ossbauer spectrum in the paramagnetic phase. It has zero trace and it is usually determined by two quantities: $V_{ZZ}$, where OZ is the principal axis and $\eta = \vert \frac {V_{YY}-V_{XX}} {V_{ZZ}} \vert$. 

The splitting $\Delta E_Q$ of the spectral doublet observed in the paramagnetic phase, due to the electric quadrupolar hyperfine interaction with the  $I$=3/2 excited nuclear state of $^{57}$Fe, with quadrupole moment $Q$=0.21\,barn, is given by:
\begin{equation}
\Delta E_Q = \frac{eQV_{ZZ}}{2} \sqrt{1+\frac{\eta^2}{3}}.
\label{deq}
\end{equation}

\begin{figure}[ht!]
\centering\includegraphics[width=0.8\linewidth]{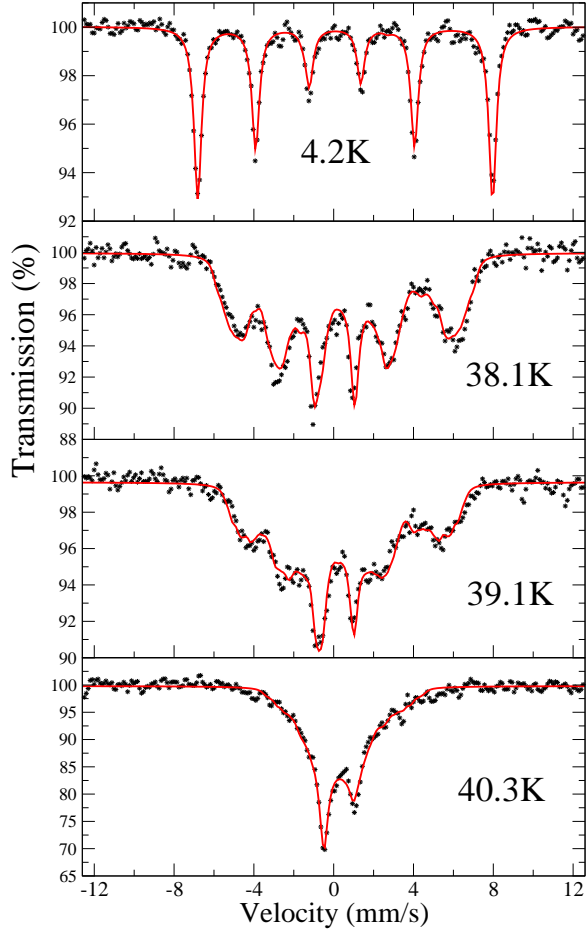}
\caption{$^{57}$Fe M\"ossbauer spectra in o-YbMnO$_3$:Fe at selected temperatures. The fits, shown by a solid red line, are as explained in the text.}
\label{spe}
\end{figure}

The experimental value $\Delta E_Q \simeq$1.54\,mm/s, obtained above 43\,K in Ref.\cite{stewart} as well as in the present work (not shown), is in rather good agreement ($\Delta E_Q \simeq$1.59\,mm/s) with the Point Charge calculation of Ref.\cite{stewart}, which also yields $\eta$=0.175. Furthermore, the ordered Mn$^{3+}$ magnetic moment lies along the orthorhombic {\bf a} axis in the low temperature E-type collinear AF phase \cite{huang}. The impurity Fe moment and the hyperfine field, proportionnal to the moment for Fe$^{3+}$, are expected to lie along {\bf a} as well. The Point Charge calculation of Ref.\cite{stewart} has determined the values of the polar and azimutal angles of the hyperfine field (the {\bf a} axis) in the Electric Field Gradient frame: $\theta \simeq$37.8$^\circ$ and $\varphi \simeq$270$^\circ$. 

\subsection{M\"ossbauer spectra in the magnetic phases}

Selected spectra are represented in Fig.\ref{spe}, in the ground collinear E-type phase at 4.2\,K, and for three temperatures inside the intermediate phase: 38.1, 39.1 and 40.3\,K. They are in good agreement with those in Ref.\cite{stewart}. 

The spectra show no significant variation in the ground E-type phase, between 4.2\,K and 36\,K, revealing the presence of a single magnetic hyperfine field, in agreement with the magnetic structure determination of a collinear single moment AF structure \cite{huang}, with moments directed along the {\bf a} axis. A good fit is obtained using the quadrupolar parameters and the polar and azimutal angles of the hyperfine field determined in Ref.\cite{stewart}. The hyperfine field at 4.2\,K is 44.3\,T, a value lying at the lower end of the typical range for Fe$^{3+}$ in insulators (50 $\pm$ 5\,T).
\begin{figure}[!ht]
\centering\includegraphics[width=0.9\linewidth,trim = 0 0 0 7cm]{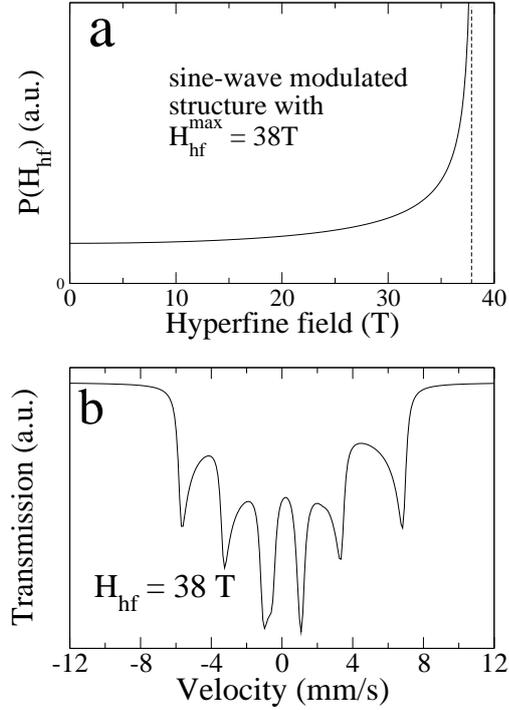}
\caption{In the presence of a sine-wave modulated collinear magnetic structure: {\bf a} Distribution of hyperfine fields at the Fe site, with a maximum value of 38\,T; {\bf b} Simulated $^{57}$Fe M\"ossbauer spectrum in the presence of this distribution.}
\label{spe_sim}
\end{figure}

Above 38\,K, in the intermediate phase, the spectra show a drastic change: the lines broaden and the resonant absorption strongly increases in the center of the spectrum, i.e. near zero velocity. At 40.3\,K, the spectrum is an almost featureless asymmetric doublet. These characteristics point to the presence of a distribution of hyperfine fields with a rather strong weight near zero and low values. Generally speaking, in a magnetically ordered phase, this very specific feature is generated in a M\"ossbauer spectrum solely by an incommensurate collinear magnetic structure, like that observed in FeVO$_4$ in the intermediate magnetic phase, for 15.7\,K $<$ T $<$ 23\,K \cite{colson}. Other types of incommensurate orderings (for instance elliptical) do not yield an enhanced weight near zero hyperfine field. 

The  distribution corresponding to a collinear incommensurate sine-wave modulation of hyperfine fields is shown in Fig.\ref{spe_sim} {\bf a}, for a maximum hyperfine field of 38\,T, and the associated simulated spectrum, with the same quadrupolar parameters and orientation of the hyperfine field as determined for o-YbMnO$_3$:Fe, in Fig.\ref{spe_sim} {\bf b}. 
Comparison of the simulated spectrum in Fig.\ref{spe_sim} {\bf b} and of the experimental spectra in the intermediate phase of o-YbMnO$_3$:Fe at 38.1\,K and 39.1\,K shows a clear similarity: not only, as mentioned above, the presence of a large spectral weight at the center of the spectrum, but also the left-hand central line being broader than the right-hand one. There is however an important difference: the outer and intermediate lines in the simulation of Fig.\ref{spe_sim} {\bf b} have an asymmetric shape, whereas those in the spectra are rather symmetrically broadened. Such a broadening could be caused by hyperfine field fluctuations, but the two other spectral features mentioned above clearly cannot, thus excluding relaxation effects as a cause for the observed peculiar spectral shapes. 

The simulated spectrum in Fig.\ref{spe_sim} {\bf b} actually corresponds to the case of a modulated magnetic structure in a pure material, where there is only one sort of magnetic ion, like in FeVO$_4$. In o-YbMnO$_3$:Fe, we observe the spectrum of Fe impurities whose presence in the matrix must entail a small perturbation of the magnetic structure of the matrix Mn ions. Furthermore, we assume that the Mn magnetic structure is locally reflected in the magnitude of the Fe impurity magnetic moment, hence of the hyperfine field at the Fe site. This assumption means that a low level substitution of the 3$d^4$ Mn$^{3+}$ ion by a 3$d^5$ Fe$^{3+}$ ion does not essentially perturb the superexchange interaction, but it is reasonable to consider that it can somehow blur the modulation of the Fe moments all over the sample.

\begin{figure}[!ht]
\centering\includegraphics[width=0.6\linewidth]{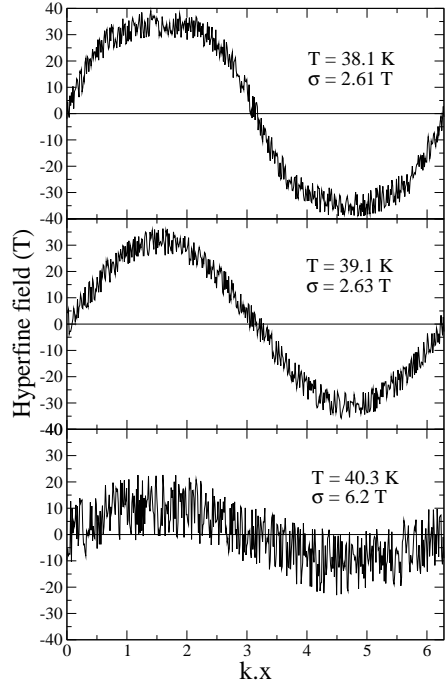}
\caption{Fitted hyperfine field (or moment) modulations in the intermediate phase of o-YbMnO$_3$:Fe at 38.1\,K, 39.1\,K and 40.3\,K showing the random deviations from the nominal modulation with their mean value $\sigma$.}
\label{mod}
\end{figure}

For these reasons, we have fitted the spectra in the intermediate phase using the four following assumptions:

i) the quadrupole interaction tensor is fixed to its value at 4.2\,K.

ii) the magnetic structure is collinear incommensurate, yielding a modulated variation for the Fe moment, and the Mn/Fe moment direction, hence that of the hyperfine field, is along the {\bf a} axis, i.e. it is the same as in the collinear E-type ground structure.

iii) the modulation is described by a Fourier expansion up to the 3 first (odd) harmonics, as a function of the abscissa $x$ along the propagation vector {\bf k}:
\begin{equation}
H_{hf}(kx) = \sum_{p=0}^2\ h_{2p+1}\ \sin[(2p+1)kx],
\end{equation}
in order to account for possible deviations from the pure sine-wave. 

iv) a small random deviation from the modulated value exists at each site, which accounts for potential defects of the incommensurate structure reflected at the impurity site. This deviation has the form: $\delta H_{\rm hf} = x \ \Delta H_{\rm hf}$, where $x$ is chosen at random in the interval [0;1] and $\Delta H_{\rm hf}$ is a parameter which must be fitted to the lineshape. The mean value of the deviation is therefore $\sigma = \frac{1}{2}\ \Delta H_{\rm hf}$.

The spectra in the intermediate phase have been successfully fitted this way, as witness the red solid lines in Fig.\ref{spe}. The corresponding hyperfine field (or moment) modulations as shown in Fig.\ref{mod}. At 38.1\,K, just above T$_{\rm N1}$ = 38.0\,K, the modulation is somewhat squared, and 3 harmonics are needed to reproduce the spectral shape: h$_1$ = 38.0\,T, h$_3$ = 4.32\,T and h$_5$ = 0.80\,T. At 39.1\,K, in the middle of the intermediate phase, the modulation is close to pure sine wave with $h_1$= 31.2\,T. At these two temperatures, the mean value of the deviation from the modulation amounts to $\simeq$6\% of the maximum hyperfine field. At 40.3\,K, just below T$_{\rm N2}$ = 41.5\,K, the first harmonics has decreased to $\simeq$11\,T and the mean value of the deviation is rather large: $\sigma \simeq$ 6.2\,T, so that the magnetic hyperfine structure has almost disappeared, leaving an asymmetric doublet with a broad base.

\section{Conclusion}

Using $^{57}$Fe M\"ossbauer spectroscopy, we have shown that the lineshapes in the intermediate magnetic phase (38.0\,K $<$ T $<$ 41.5\,K) of orthorhombic YbMnO$_3$ (doped with 2\% $^{57}$Fe) are compatible with a collinear incommensurate magnetic structure. The Fe hyperfine field, and hence the Mn spontaneaous moment, has the same direction as in the ground E-type AF phase, i.e. the crystal {\bf a} axis. The modulation is mainly sine-wave, but we could detect some ``squaring'' just above the lock-in transition. Since this type of magnetic structure has been found for the intermediate phase in orthorhombic RMnO$_3$ with R = Gd-Ho \cite{goto,kimura,kenzelmann,arima,munoz}, we think our M\"ossbauer spectra demonstrate its presence also in orthorhombic YbMnO$_3$. Since M\"ossbauer spectroscopy is a local microscopic technique, it cannot determine the wave vector of the modulation, and this should be done by neutron diffraction.

\section*{Acknowledgments}
We thank the ``Service Synth\`eses Hautes Pressions'' of the ICMCB (Bordeaux) for the preparation of the orthorhombic YbMnO$_3$ sample.

\end{document}